# Exchange striction induced thermal Hall effect in van der Waals antiferromagnet MnPS$_3$

Heejun Yang,[1] Gyungchoon Go,[2] Jaena Park,[1] Se Kwon Kim,[2] and Je-Geun Park[1,*]

[1]Department of Physics and Astronomy, Seoul National University, Seoul 08826, Korea

[2]Department of Physics, Korea Advanced Institute of Science and Technology, Daejeon 34141, Korea

[*] Corresponding author: jgpark10@snu.ac.kr

## Abstract

The thermal Hall effect has emerged as an ideal probe for investigating topological phenomena of charge-neutral excitation. Notably, it reveals crucial aspects of spin-lattice couplings that have been difficult to access for decades. However, the exchange striction mechanism from a lattice-induced change in exchange interaction has often been ignored in thermal Hall experiments. MnPS$_3$ can offer a platform to study exchange striction on the thermal Hall effect due to its significant spin-lattice coupling and field-induced non-collinear spin configuration. Our thermal transport data show distinct temperature and field dependence of longitudinal thermal conductivity ($\kappa_{xx}$) and thermal Hall effect ($\kappa_{xy}$). By using detailed theoretical calculation, we found that the inclusion of the exchange striction is essential for a better description of both $\kappa_{xx}$ and $\kappa_{xy}$. Our result demonstrates the importance of the exchange striction mechanism for a complete understanding of the magnon-phonon-driven thermal Hall effect.

# Ⅰ. INTRODUCTION

The topological nature of spin excitations has recently become the focus of recent extensive studies, both experiments and theories, whose potential consequences go beyond the horizon of fundamental understanding and spintronics [1,2]. Due to their charge neutrality, the thermal Hall effect (THE) is considered a promising experimental technique to investigate this exotic aspect of magnons, otherwise fundamental quasiparticles of magnetism [3]. Early measurements of THE have been reported in various magnetic insulators: ferromagnetic spin configurations [4–7] and frustrated magnets [8–17]. However, recent unexpected THE observed in non-magnetic materials [18–21] demonstrate the potentially important role of phonons in THE studies. Another closely related recent report is that strong spin fluctuations can significantly modify thermal transport in the paramagnetic phase of magnetic insulators like hexagonal manganite $YMnO_3$ [22]. This new development requires one to put spins and phonons on equal footing, which poses a severe, experimental and theoretical, challenge to our understanding of heat transport, including THE.

Therefore, spin-lattice coupling has become an urgent and timely issue in understanding thermal transport in general and THE specifically. For example, recent experimental works, including cuprates, proposed a significant role of spin-lattice coupling for THE [23–32]. Here, one has to note that there are two fundamental mechanisms for spin-lattice coupling. One is the single-ion magnetostriction type of spin-lattice coupling that arises from the modulation of crystal fields surrounding magnetic ions [33–36]. Incidentally, it has been suggested as the source of several THE experiments [7,37,38]. Another equally important mechanism is exchange striction, which is induced by a spatial modulation of magnetic ions changing the strength of exchange interaction [39]. In particular, the exchange striction of Heisenberg interaction could be applicable to a wide range of magnetic systems, since the spin-orbit coupling is not, in principle, be required for this mechanism [40]. Indeed, several inelastic neutron and X-ray studies have been done to realize the importance of the exchange striction [41–47]. Several theoretical studies of THE based on exchange striction have been suggested, accordingly [48,49]. Nevertheless, experimentally, it is still an unanswered question how the exchange striction mechanism produces THE for a given system, begging for comprehensive studies in the context of THE data analysis.

Here, two-dimensional (2D) layered magnets, known as van der Waals magnets, seem to be an ideal test ground with strong spin-lattice coupling. In particular, antiferromagnetic $TMPS_3$ (TM = Mn, Fe, and Ni) has been reported to have highly enhanced spin-lattice coupling [50–56]. As a member of $TMPS_3$ materials, $MnPS_3$ is an ideal material for investigating the exchange striction affecting THE for the following reasons. First, $MnPS_3$ consists of 2D honeycomb layers of $Mn^{2+}$ ions [Fig. 1(a)], with a Néel-type antiferromagnetic ground state dominated by Heisenberg interaction [57,58]. As two opposite directions of magnon THE cancel out in for the Heisenberg model [59], it is natural to consider spin-lattice coupling for finite THE in $MnPS_3$ [36,60]. Next, Raman studies on $MnPS_3$ observed that the 155 $cm^{-1}$ peak shifts drastically at the Néel temperature ($T_N$) [50–52], indicating significant spin-lattice coupling. Lastly, the magnetic ground state of $MnPS_3$ involves $Mn^{2+}$ moments tilted away from the *z*-axis (out-of-plane direction) at an angle of 8° [61]. Due to this tilting, a non-collinear spin configuration can be induced by applying the magnetic field along the *z*-axis, as described in the right side of Fig. 1(a). Notably, this non-collinearity is necessary for magnon-phonon hybridization via the exchange striction by Heisenberg interaction [39]. All three considerations put $MnPS_3$ in a unique position for the THE study regarding exchange striction.

This study presents thermal transport measurement data of $MnPS_3$, a 2D honeycomb Néel-type antiferromagnet. Our experimental results revealed the distinct field dependence of longitudinal thermal conductivity ($\kappa_{xx}$) and finite thermal Hall conductivity ($\kappa_{xy}$). To analyze the experimental data, we constructed a model Hamiltonian with terms included for both types of spin-lattice coupling: single-ion magnetostriction and exchange striction. Based on these comprehensive studies, we conclude that both exchange striction and single-ion magnetostriction play distinct roles in the magnon-phonon-driven THE of MnPS$_3$.

## II. EXPERIMENTAL METHODS

High-quality single crystals of MnPS$_3$ were grown by the chemical vapor transport method. Manganese powder (99.95%, Alfa Aesar), red phosphorus (≥99.99%, Sigma-Aldrich), and sulfur powder (99.98%, Sigma-Aldrich) were mixed in the stoichiometric ratio with 5 wt% of extra sulfur within an argon atmosphere [62,63]. The mixture was sealed into a quartz ampule, and the quartz ampule was put into a horizontal two-zone furnace with a temperature difference of 780 and 700 °C for seven days. Then, the quartz ampule was cooled to room temperature over two days. Several green plate-like hexagonal-shaped crystals of MnPS$_3$ were obtained from the quartz ampule. The typical size of the crystals was 3×3×0.1 mm$^3$. Energy-dispersive X-ray (EDX) spectroscopy (Quantax 100, Bruker & EM-30, Coxem) confirmed the stoichiometry of the sample as Mn:P:S = 0.988:0:991:3.021. The magnetization ($M$) of MnPS$_3$ sample was measured using MPMS-XL5 (Quantum Design) by applying the magnetic field along the $z(c^*)$-axis.

Thermal transport measurement was performed by the standard steady-state method using one heater and three thermometers. In this study, we employed custom-made SrTiO$_3$ parallel plate capacitors as thermometers to minimize errors from high magnetic fields and performed *in situ* calibrations [64]. The heat current was applied along the $xy(ab)$-plane, and the magnetic field was applied along the $z(c^*)$-axis. Three thermometers measured longitudinal ($\Delta T_x$) and transverse ($\Delta T_y$) temperature differences simultaneously. Under the isothermal condition, the field dependence of $\Delta T_x$ data was collected with a slowly varying field sweep. On the other hand, for $\Delta T_y$ measurement, a static magnetic field was applied in a step-by-step mode, and every data point was taken by averaging $\Delta T_y$ for a few minutes to avoid error possibly coming from the magnetocaloric effect. After that, we antisymmetrized $\Delta T_y$ data with respect to the magnetic fields to eliminate longitudinal contamination in $\Delta T_y$ due to contact misalignment. Resulting $\Delta T_x$ and antisymmetrized $\Delta T_y$ data were converted into $\kappa_{xx}$ and $\kappa_{xy}$ respectively, using Fourier's law of heat conduction. Further experimental details are presented in the Supplemental Material [65].

## III. RESULTS

As seen in Fig. 1(b), the magnetization ($M$) data of our MnPS$_3$ sample reproduced several key features found in the previous report [71]. Firstly, the temperature dependence of $M$ shows a broad maximum at around 120 K, implying a short-range spin-spin correlation [72]. Next, we can find a sharp cusp at 79 K [see the inset of Fig. 1(b)], which indicates $T_N$. Both field cooling and zero-field cooling measurements showed no discernible difference, consistent with previous study [73]. Note that our $M$ data showed a negligibly small Curie tail in the low-temperature range, which supports the high quality of our MnPS$_3$ crystals.

As shown in Fig. 1(c), the temperature dependence of $\kappa_{xx}$ exhibits general features of phonon thermal conductivity with a single peak around 20 K and a monotonic decrease for $T$>20 K [74]. When applying a field of 9 T, we observed apparent suppression of $\kappa_{xx}$ at low temperatures below 50 K. Although $\kappa_{xx}$ seems to display a smooth change as the temperature crosses $T_N$, we can also observe a small cusp around $T_N$ [see the inset of Fig. 1(c)]. This small anomaly is quite consistent with the Raman studies reporting drastic shifts of 155 cm$^{-1}$ peak at $T_N$, indicating significant spin-lattice coupling in MnPS$_3$ [50–52].

Fig. 1(d) also shows $\kappa_{xy}$ as a function of temperature with a negative peak around 20 K. However, its size rapidly diminishes as the temperature approaches $T_N$, exhibiting a cusp instead of the nearly smooth behavior observed in $\kappa_{xx}$. For $T>T_N$, the temperature dependence of $\kappa_{xy}$ is much weaker than in the magnetically ordered phase, representing almost flat $\kappa_{xy}$ up to 200 K. These observations lead us to assume that the long-range magnetic ordering is crucial for finite $\kappa_{xy}$ in MnPS$_3$. The small remnant $\kappa_{xy}$ signals in the supposedly paramagnetic phase above $T_N$ are likely to be due to short-range fluctuations as proposed by recent THE studies [75,76].

As seen in Fig. 2(a), we present the magnetization ($M$) data as a function of the magnetic field obtained at various temperature points. For $T \ll T_N$, a spin-flop transition exhibits a sharp step-like increase in $M$ around 4 T. As the temperature is increased further, the step-like increase of $M$ becomes more moderate, eventually turning into a linear behavior with respect to the magnetic field. Consequently, it becomes harder to identify the spin-flop transition above 50 K from the $M$ measurement alone. This field-linear behavior of $M$ has been maintained up to 200 K.

Fig. 2(b) shows the detailed field dependence of magneto-thermal conductivity defined as $\Delta\kappa_{xx}(H)/\kappa_{xx}(0) \equiv [\kappa_{xx}(H) - \kappa_{xx}(0)]/\kappa_{xx}(0)$. The complex behavior can be divided into three temperature ranges. First, $\kappa_{xx}$ gets increased slightly at the low field range for $T \ll T_N$ before falling sharply during the spin-flop transition. Next, for $0.5T_N<T<1.5T_N$, $\kappa_{xx}$ decreases even at the low field range. In addition, we can only see a small cusp around the spin-flop transition rather than a large suppression. Note that this cusp becomes less distinct as the temperature increases and is absent for $T>T_N$. Finally, for $T>1.5T_N$, $\kappa_{xx}$ increases again across the entire field range.

Fig. 2(c) displays field dependence of $\kappa_{xy}$ data. For $T \ll T_N$, we could only obtain noisy negative $\kappa_{xy}$ data exhibiting monotonic field response, due to the small signal size and the resolution limit of our experimental setup [64]. However, at slightly higher temperatures, we could observe a hump around 4.5 T in addition to the negative background. This hump quickly diminishes as the temperature increases. For $T>0.5T_N$, it becomes challenging to identify the hump, and $\kappa_{xy}$ becomes field-linear above $T_N$.

## IV. DISCUSSIONS

### A. Debye-Callaway model fitting for $\kappa_{xx}$ data

The temperature dependence of $\kappa_{xx}$ can be characterized by various scattering mechanisms of heat carriers [74]. For the phonon-dominant $\kappa_{xx}$ case, the Debye-Callaway model is usually a standard starting point for figuring out the detailed phonon scattering mechanisms [77]. With the Debye-Callaway model, the phonon thermal conductivity can be written as follows:

$$\kappa_{xx} = \frac{k_B^4}{2\pi^2 v_D \hbar^3} T^3 \int_0^{T_D/T} \frac{x^4 \exp(x)}{[\exp(x)-1]^2} \tau(\omega,T)\, dx, \qquad (1)$$

where $k_B$ ($\hbar$) is the Boltzmann (the reduced Planck) constant, $T_D$ ($v_D$) is the Debye temperature (the group velocity of acoustic phonons) estimated from the Debye model, $\tau^{-1}(\omega,T)$ is the scattering rate of the phonon, $\omega$ is the frequency of the phonon, and $x = \hbar\omega/k_B T$. According to the Debye model, the relationship between $T_D$ and $v_D$ is given as $T_D = v_D \frac{\hbar}{k_B}(6\pi^2 n)^{1/3}$, where $n$ is the number of atoms per unit volume. For MnPS$_3$, $T_D$ has been estimated at 177 K [78].

Representative scattering sources for phonon heat conduction can be listed as follows [74]: sample boundary ($\tau_{BD}^{-1} = v_D/d$), linear defects ($\tau_{LD}^{-1} = A_0 \omega$), point defects ($\tau_{PD}^{-1} = A_1 \omega^4$), and Umklapp process ($\tau_U^{-1} = A_2 \omega^2 T \exp(-T_D/bT)$), where $d$ represents a typical dimension of the system, $b$ is the characteristic constant for the Umklapp process, and $A_0$, $A_1$ and $A_2$ are free parameters. Then, $\tau^{-1}(\omega, T)$ can be approximated by a sum of possible scattering sources following Matthiessen's rule,

$$\tau^{-1}(\omega, T) = \tau_{BD}^{-1} + \tau_{LD}^{-1} + \tau_{PD}^{-1} + \tau_U^{-1}. \qquad (2)$$

To minimize the model, we fixed $d$ as 1 mm as the typical sample size and used conventional value for $b = 2\sqrt[3]{N} \sim 5.43$, where $N$ is the number of atoms in a unit cell [22].

Hence, we tried to fit the experimental $\kappa_{xx}$ data taken with 0 T using three free parameters $A_0$, $A_1$ and $A_2$ via least-square criterion. Our best-fit result can be seen as the black solid curve in Fig. 1(c). For the fitting procedure, we excluded the data points around $T_N$; we used the data points for $T<0.5T_N$ and $T>1.5T_N$. The main reason is that the fitting quality worsened when we fully included the data points around $T_N$. In the end, we obtained the following parameter set $A_0 = 4.32 \times 10^{-5}$, $A_1 = 3.99 \times 10^{-44}$ s$^3$, and $A_2 = 4.10 \times 10^{-18}$ s/K, which are comparable to those values from recent studies employing the Debye-Callaway model [22,29,79]. The experimental data was quite consistent with the fitting result around 20 K. However, we also note that the $\kappa_{xx}$ data decreases faster than the fitting curve above 40 K.

As shown in Fig. 1(c), the suppressed behavior of $\kappa_{xx}$ data as compared to the Debye-Callaway model of Eq. (2) demonstrates the existence of additional phonon scattering sources. Interestingly, a recent inelastic neutron study reported that magnetic excitations of MnPS$_3$ persist up to 200 K, indicating robust short-range spin-spin correlations even above $T_N$ [80]. Therefore, phonon scattering by spin fluctuations could also be considered since the suppression of $\kappa_{xx}$ was severe around the $T_N$, implying significant spin-lattice coupling in MnPS$_3$ [22]. Although we can achieve better $\kappa_{xx}$ fitting by obtaining detailed information on spin-lattice coupling via *ab-initio* calculation [81,82], it would be beyond the scope of our study, calling for future works.

## B. Correlation between spin-flop transitions and thermal transport coefficients

As displayed in Figs. 2(b) and (c), the field dependences of both $\kappa_{xx}$ and $\kappa_{xy}$ show some anomalies around 4.5 T, which seems related to the spin-flop transition. For a detailed comparison between the spin-flop transition and the anomalies in $\kappa_{xx}$ and $\kappa_{xy}$, we first provide the contour plot of $dM/dH$, the derivative of $M$ with respect to magnetic field [see Fig. 3(a)]. We could define the spin-flop transition by choosing the field point for the maximum value of $dM/dH$ at each temperature. With this definition, the spin-flop transition happens around 4 T at the lowest temperature. When the temperature increases, we can observe that the spin-flop transition occurs at slightly larger magnetic fields. Above 60 K, we could not resolve the spin-flop transition via $M$ measurement due to the limited field range of our magnetometer.

Next, Fig. 3(b) provides the contour plot of $d\kappa_{xx}/dH$, the field-derivative of $\kappa_{xx}$. The blue colored area represents anomalies of $\kappa_{xx}$, i.e., sharp suppression. We also plotted field points for the maximum value of both $dM/dH$ and $|d\kappa_{xx}/dH|$ together on Fig. 3(b). Interestingly, temperature dependences of these field points are comparable to each other, indicating a close correlation between the spin-flop transition and anomalies of $\kappa_{xx}$. Unfortunately, the contour plot of $d\kappa_{xy}/dH$ could not be provided due to coarse data points. Nevertheless, we can repeatedly observe the hump of $\kappa_{xy}$ at 4.5 T for the low-temperature region [see Fig. 2(c)]. Since the spin-flop transition happens around 4.5 T below 50 K, we could conclude that the anomalies of $\kappa_{xy}$ are also closely related to the spin-flop transition.

## C. Linear spin-wave theory for $\kappa_{xy}$ data

Berry curvature has been regarded as a primary source of THEs via charge-neutral excitations, since non-zero Berry curvature can induce additional transverse motion of wave packets analogous to anomalous Hall effect of electronic system [83]. Hence, we tried analyzing our data via Berry curvature scenario first. Considering all the realistic scenarios, magnons could be a main candidate for THE for the magnetically ordered phase. For Néel-type honeycomb antiferromagnet, it is known that non-zero bond-dependent exchange interactions are essential for finite $\kappa_{xy}$ [84]. However, as reported in several magnetic insulators, the size of $\kappa_{xy}$ due to magnons often increases as the temperature rises and collapses quickly near the magnetic phase transition [4,5,7]. This is because the nontrivial Berry curvature is usually concentrated around the Brillouin zone (BZ) boundary [85–87], where the energy scale is similar to $T_N$. In MnPS$_3$ the magnon energy at the BZ boundary is approximately 11 meV [57]. Therefore, the resulting $\kappa_{xy}$ would increase up to $T_N$, which is inconsistent with our experimental data.

Another candidate worth considering is the magnon-phonon hybrid excitation [34–36,48,49,88]. When magnon and phonon modes cross each other, the non-zero spin-lattice coupling could open a gap at the crossing point, resulting in magnon-phonon hybridizations and the finite Berry curvature [see Fig. 4(a)]. In most cases, the coupling between the acoustic phonon branch and low-lying magnon near the BZ center would mainly affect the low-temperature range of $\kappa_{xy}$. For instance, the latest study of VI$_3$ explained the enlarged size of $\kappa_{xy}$ around 20 K by including the spin-lattice coupling after finding that a magnon-only model cannot explain the experimental data [7]. As our $\kappa_{xy}$ also exhibits a prominent peak around 20 K, it is reasonable to expect that magnon-phonon hybridization could play a significant role for THE in MnPS$_3$.

To analyze the experimental results, we employed linear spin-wave theory (LSWT) and calculated magnon-phonon hybridization numerically. To simplify the analysis, we assumed a perfect honeycomb lattice of Mn$^{2+}$ ions despite very tiny distortion on honeycomb structure of actual MnPS$_3$ system. We also carried out our calculations for the monolayer limit, which can be justified by the negligibly small interlayer coupling [57].

Our magnetic Hamiltonian ($\mathcal{H}_m$) consists of Heisenberg interactions ($J_n$), easy-axis anisotropy ($D$), and Zeeman term,

$$\mathcal{H}_m = \sum_{\langle ij \rangle_n} J_n \mathbf{S}_i \cdot \mathbf{S}_j + D \sum_i (\hat{\mathbf{n}} \cdot \mathbf{S}_i)^2 - g\mu_B\mu_0 H \sum_i S_i^z, \tag{3}$$

where $\mathbf{S}_i$ is the spin vector of the $i$-th site, $\hat{\mathbf{n}} = \sin 8°\, \hat{\mathbf{x}} + \cos 8°\, \hat{\mathbf{z}}$ is a unit vector necessary to reproduce the ground state of MnPS$_3$ exhibiting 8° of canting from the $z$-axis. $g$, $\mu_B$, $\mu_0$, and $H$ denote $g$-factor, Bohr magneton, vacuum permeability, and external magnetic field applied along the $z$-axis, respectively. Throughout this study, we put $x(z)$-axis parallel to the crystallographic $a(c^*)$-axis [see Fig. 1(a)]. We used $S = 5/2$, $J_1 = 1.54$ meV, $J_2 = 0.14$ meV, $J_3 = 0.36$ meV, and $D = -0.00215$ meV as taken from the previous reports [58,59], and $g = 2$ from recent electron spin resonance study [89]. The spin configurations under the finite magnetic field applied along the $z$-axis were obtained through classical energy minimization.

Next, the phonon Hamiltonian ($\mathcal{H}_p$) is

$$\mathcal{H}_{\mathrm{p}} = \sum_i \frac{\mathbf{p}_i^T \mathbf{p}_i}{2m} + \frac{1}{2}\sum_{i,j} \mathbf{u}_i^T K(\mathbf{R}_i - \mathbf{R}_j)\mathbf{u}_j, \qquad (4)$$

where $\mathbf{u}_i(\mathbf{p}_i)$ is the displacement (conjugate momentum) vector of $i$-th $Mn^{2+}$ ion and $K$ is a spring constant. To simplify the model, we ignored non-magnetic ions and only considered $K$ for the 1st nearest-neighbor bonds. Hence, $\mathbf{R}_i$ is equilibrium position of $i$-th $Mn^{2+}$ ion, and $K(\mathbf{R}_i - \mathbf{R}_j)$ is a spring constant between $i$-th and $j$-th $Mn^{2+}$ ions. Note that, by choosing suitable $K$, our acoustic phonon branch is comparable to the result of a recent *ab-initio* study [65,90].

For the spin-lattice coupling, we considered two mechanisms for our minimum model, as discussed in the introduction: single-ion magnetostriction ($\mathcal{H}_{\mathrm{mp,SI}}$) and exchange striction ($\mathcal{H}_{\mathrm{mp,ex}}$) [39].

$\mathcal{H}_{\mathrm{mp,SI}}$ reads in the hexagonal lattice [91]:

$$\mathcal{H}_{\mathrm{mp,SI}} = B_1 \sum_i \left[ (S_i^x)^2 \varepsilon_i^{xx} + (S_i^y)^2 \varepsilon_i^{yy} + (S_i^x S_i^y + S_i^y S_i^x)\varepsilon_i^{xy} \right] + B_2 \sum_i \left[ (S_i^z)^2 (\varepsilon_i^{xx} + \varepsilon_i^{yy}) \right]$$

$$+ B_3 \sum_i \left[ (S_i^y S_i^z + S_i^z S_i^y)\varepsilon_i^{yz} + (S_i^x S_i^z + S_i^z S_i^x)\varepsilon_i^{xz} \right], \qquad (5)$$

where $\varepsilon_i^{\mu\nu}$ is a one-ion strain [65,92], $B_1 = -\frac{B^\gamma}{2}$, $B_2 = -\frac{B^\gamma}{4} - B^\alpha$, and $B_3 = -\frac{B^\epsilon}{2}$ with $B^\alpha = \frac{\sqrt{3}}{2} B^{\alpha,1,2} - \frac{1}{4} B^{\alpha,2,2}$. Here, the coupling constant $B^{\alpha,1,2}$ represents the volume change, $B^{\alpha,2,2}$ is for change in the ratio between the crystallographic *a*- and *c*-axes, $B^\gamma$ is for shear motion in the *ab*-plane, and $B^\epsilon$ is for shear motion in planes, including the *c\**-axis [93]. Note that in LSWT $B_3$ only couples to the out-of-plane vibrational motion while $B_1$ and $B_2$ only do to in-plane motions.

We can now derive $\mathcal{H}_{\mathrm{mp,ex}}$ by performing the Taylor-series expansion of the exchange interaction as a function of the interatom distance vector [39]. In principle, various types of exchange interactions can be expanded by Taylor-series resulting in spin-lattice coupling, such as Heisenberg, Dzyaloshinskii-Moriya, or Kitaev-like interactions. However, it is expected that $\mathcal{H}_{\mathrm{mp,ex}}$ of Heisenberg interaction would be dominant for $MnPS_3$ due to high-spin configuration of $Mn^{2+}$ ion which possesses weak spin-orbit coupling effects [81]. Therefore, we only considered the case of $J_1$ for simplicity as follows:

$$\mathcal{H}_{\mathrm{mp,ex}} = J_1 \alpha_{\mathrm{ex}} \sum_{\langle ij \rangle_1} \left[ (\mathbf{u}_i - \mathbf{u}_j) \cdot \frac{\mathbf{r}_{ij}}{|\mathbf{r}_{ij}|^2} \right] (\mathbf{S}_i \cdot \mathbf{S}_j), \qquad (6)$$

where $\mathbf{r}_{ij}$ is the vector connecting two 1st nearest-neighbor $Mn^{2+}$ ions and $\alpha_{\mathrm{ex}}$ is a dimensionless coupling constant that can be obtained from the pressure dependence of 1st nearest-neighbor bond length and $T_N$ using the following formula [41]: $\alpha_{\mathrm{ex}} = \frac{|\mathbf{r}_{ij}|(P_0)}{T_N(P_0)} \cdot \frac{(\partial T_N/\partial P)}{(\partial |\mathbf{r}_{ij}|/\partial P)}$. As we are assuming a monolayer of $Mn^{2+}$ ions, $\mathcal{H}_{\mathrm{mp,ex}}$ only describes the coupling between spins and in-plane vibration. Therefore, in-plane uniaxial pressure ($P_{ab}$) study is required for estimating $\alpha_{\mathrm{ex}}$ in our case. Unfortunately, only the hydrostatic pressure ($P_h$) studies have been reported for $MnPS_3$ [94,95] giving $\alpha_{\mathrm{ex}} \sim -47$. Interestingly, the previous observation that $\partial T_N/\partial P_{ab}$ is several times larger than $\partial T_N/\partial P_h$ in $Cr_2Ge_2Te_6$, another 2D van der Waals magnet [96], we think, justifies our estimate of $|\alpha_{\mathrm{ex}}|$ for $MnPS_3$.

By diagonalizing the total Hamiltonian $\mathcal{H}_{\mathrm{tot}} = \mathcal{H}_{\mathrm{m}} + \mathcal{H}_{\mathrm{p}} + \mathcal{H}_{\mathrm{mp,SI}} + \mathcal{H}_{\mathrm{mp,ex}}$, we can obtain the band dispersion and Berry curvature. We then calculated $\kappa_{xy}$ using the following formula [83]:

$$\kappa_{xy} = -\frac{k_B^2 T}{\hbar(2\pi)^2 d_c} \sum_{\mathbf{k}} \sum_{\lambda=1}^{N_{\text{tot}}} c_2[\rho(E_{\lambda,\mathbf{k}})]\Omega_{\lambda,\mathbf{k}}, \qquad (7)$$

where $d_c = 6.49$ Å is the thickness of the honeycomb layer [97], $N_{\text{tot}}$ is the number of total bands, $\Omega_{\lambda,\mathbf{k}}$ is the Berry curvature of the $\lambda$-th energy level at $\mathbf{k}$-point $E_{\lambda,\mathbf{k}}$, and $c_2[\rho(E_{\lambda,\mathbf{k}})]$ is the weight function of the Bose-Einstein distribution $\rho(E_{\lambda,\mathbf{k}}) = [\exp(E_{\lambda,\mathbf{k}}/k_B T) - 1]^{-1}$. During the calculation, the temperature is reduced by a mean-field calculation of $T_N = \frac{1}{3k_B} S(S+1)(3J_1 - 6J_2 + 3J_3)$ [57]. To identify the role of each spin-lattice coupling constant clearly, we considered only the out-of-plane vibration for $B^\epsilon$ ($N_{\text{tot}} = 4$) and in-plane vibration for other coupling constants ($N_{\text{tot}} = 6$). We present further details in the Supplemental Material [65].

First, we calculated $\kappa_{xy}$ as a function of the magnetic field by including each coupling constant separately [Fig. 4(b)]. In our theoretical studies, finite $\kappa_{xy}$ was only found for the $B^\gamma$ and $B^\epsilon$ cases, which shows an anomaly around the spin-flop transition. Upon closer inspection, the $B^\gamma$ term reproduces the hump shape as in the experimental data rather than the pulse shape given by $B^\epsilon$. This implies that in-plane vibration is essential for THE in MnPS$_3$, which is consistent with the recent Raman study [52]. However, note that the monotonic negative background of the experimental THE data could not be reproduced by $B^\gamma$ only case. Below, the inclusion of the exchange striction will be shown to reproduce the monotonic negative background.

Next, we checked how $\kappa_{xy}$ changes when two different kinds of coupling constants are included simultaneously. For simplicity, we considered only the in-plane vibration since $B^\gamma$ is essential to reproduce the hump shape of $\kappa_{xy}$ at the spin-flop transition. Firstly, we found that combining $B^\gamma$ and $B^\alpha$ only changed the size and sign of $\kappa_{xy}$ at the spin-flop transition [see Fig. 4(c)]. For instance, when $B^\gamma$ and $B^\alpha$ have the same sign, the anomaly of $\kappa_{xy}$ becomes increased in the negative direction. On the contrary, when $B^\gamma$ and $B^\alpha$ have the opposite sign each other, the anomaly of $\kappa_{xy}$ becomes enhanced in the positive direction. Unfortunately, the negative background of $\kappa_{xy}$ could not be reproduced yet.

Interestingly, when combining $B^\gamma$ and $\alpha_{\text{ex}}$, we can reproduce the monotonic background, as shown in Fig. 4(d), in addition to the hump. To be more specific, the negative background of $\kappa_{xy}$ was obtained when the signs of the $B^\gamma$ and $\alpha_{\text{ex}}$ are the same. On the other hand, when the signs of the $B^\gamma$ and $\alpha_{\text{ex}}$ are opposite each other, the positive background prevails. Note that the strength of monotonic background is proportional to the size of $\alpha_{\text{ex}}$. From this, we can conclude that $\mathcal{H}_{\text{mp,SI}}$ and $\mathcal{H}_{\text{mp,ex}}$ are crucial to reproduce the hump at the spin-flop transition and the negative background of $\kappa_{xy}$, respectively.

Lastly, by choosing the appropriate size of parameters ($B^\gamma = -0.04$ meV/Å$^2$, $B^\alpha = -0.25 B^\gamma$, and $\alpha_{\text{ex}} = -3 \cdot 47$) comparable to the previous studies [35,36,39,60], we also calculated $\kappa_{xy}$ at different temperature points [Fig. 4(e)]. Interestingly, the computational results qualitatively reproduce our experimental data well: with increasing temperature, the relative size of the hump around the spin-flop transition becomes smaller compared to the size of the negative background.

To further investigate the role of $\alpha_{\text{ex}}$ to achieve the negative background of $\kappa_{xy}$ when combining $B^\gamma$ and $\alpha_{\text{ex}}$, we compared the band dispersion and Berry curvature for two different cases at the magnetic field of 9 T in Fig. 5. When displaying the band dispersion, we employed the parameter $p_{\lambda,\mathbf{k}}^{\text{ph}}$ which indicates the degree of the phonon nature of the $\lambda$-th mode at the momentum $\mathbf{k}$, e.g., $p_{\lambda,\mathbf{k}}^{\text{ph}} = 1$ for a pure phonon, $p_{\lambda,\mathbf{k}}^{\text{ph}} = 0$ for a pure magnon, and $0 < p_{\lambda,\mathbf{k}}^{\text{ph}} < 1$ for hybridized modes [45]. In this study, we have assigned $\lambda$ to each mode in descending order in $E_{\lambda,\mathbf{k}}$. On the other hand, we only provided the Berry curvature of the lowest energy band since THE is mainly characterized by the lowest band for the the low-temperature range.

The first case is the $B^\gamma$ only model. As expected, magnon-phonon hybridization occurs at magnon-phonon crossing points [see Fig. 5(a)]. The resulting Berry curvature is shown in Fig. 5(c), in which the Berry curvature is concentrated along a star-shape with six corners. The next case is for combining $B^\gamma$ and $\alpha_{\text{ex}}$ with the same sign. As seen in Fig. 5(b), the magnon-phonon hybridization becomes stronger, showing a clear visible gap. The resulting Berry curvature is given in Fig. 5(d). The most obvious difference compared to the $B^\gamma$ only case is that the Berry curvature hot-spots can be seen at each corner of the star shape. Note that four of them are positively signed, and the others are negatively signed. According to Eq. (7), $\kappa_{xy}$ is negatively proportional to the Berry curvature. Hence, it is natural to consider that the negative background of $\kappa_{xy}$ would come from positively signed Berry curvature hot-spots.

Despite a successful description of the overall field dependence of $\kappa_{xy}$, we acknowledge that there is still a discrepancy at a quantitative level. Note that a similar discrepancy was also found in other recent THE studies [7,38]. We think that this discrepancy could be mainly due to the oversimplified $\mathcal{H}_{\text{p}}$ used for our theoretical studies. In particular, considering phonons due to all non-magnetic ions like P and S of MnPS$_3$ would be an important but demanding natural next step forward, as demonstrated for a couple of triangular systems [39,41,45].

Another way to improve the data analysis is to consider other mechanisms. Firstly, we can consider whether the THE mechanism of non-magnetic SrTiO$_3$ applies to MnPS$_3$, since a small remnant $\kappa_{xy}$ signal persists in the paramagnetic phase up to 200 K. However, it is known that a huge dielectric constant (~$10^4$) is essential for THE of SrTiO$_3$ [98], which is not the case with MnPS$_3$ [99]. On the other hand, magnetic dipolar interaction could be an additional channel of spin-lattice coupling, resulting in Berry curvature [88]. In addition, phonon skew-scattering by magnetic excitations or defects can be an alternative scenario, as suggested by the latest THE theories [76,100,101]. Advanced computational techniques such as nonlinear spin-wave theory or *ab-initio* calculation would be required [81,82,102], and all these go beyond the scope of the current work and will be the subject of future studies.

### D. Estimating field dependence of phonon dominant $\kappa_{xx}$

In the last part, we tried to investigate the field dependence of $\kappa_{xx}$ by using the Boltzmann transport equation [74] with magnon-phonon hybridized bands

$$\kappa_{xx} \propto \sum_{\mathbf{k}} \sum_{\lambda=1}^{N_{\text{tot}}} \left[ E_{\lambda,\mathbf{k}} \frac{\partial \rho(E_{\lambda,\mathbf{k}})}{\partial T} \right] v_{\lambda,\mathbf{k}}^2 \tau_{\lambda,\mathbf{k}}, \qquad (8)$$

where $v_{\lambda,\mathbf{k}}$ and $\tau_{\lambda,\mathbf{k}}$ are the group velocity and relaxation time, respectively. The factors $\left[ E_{\lambda,\mathbf{k}} \frac{\partial \rho(E_{\lambda,\mathbf{k}})}{\partial T} \right] v_{\lambda,\mathbf{k}}^2$ and $\tau_{\lambda,\mathbf{k}}$ in Eq. (8) are mainly influenced by detailed band structure and scattering processes, respectively. In principle, below $T_{\text{N}}$, $\kappa_{xx}$ includes the phonon ($\kappa_{xx}^{\text{ph}}$) and magnon ($\kappa_{xx}^{\text{mag}}$) contributions. Nevertheless, we would like to assume that $\kappa_{xx}^{\mathrm{mag}}$ is negligible in the case of MnPS$_3$, since the $\kappa_{xx}(T)$ data of Fig. 1(c) already show representative phonon behavior and magnon heat transport often requires significant exchange interactions ($J \sim 1500$ K) [103,104].

To estimate phonon contribution ($\kappa_{xx}^{\mathrm{ph}}$) from magnon-phonon hybridized excitations, we tried a similar approach to that introduced in the recent thermal transport study [105]: we multiplied $p_{\lambda,\mathbf{k}}^{\mathrm{ph}}$ to Eq. (8). For details see the Supplemental Material [65]. In general, $\tau_{\lambda,\mathbf{k}}$ is determined by various scattering sources such as sample boundary, point defect, and others [74]. Here we assumed $\tau_{\lambda,\mathbf{k}}$ to be constant for simplicity and calculated $\Delta\kappa_{xx}^{\mathrm{ph}}(H)/\kappa_{xx}^{\mathrm{ph}}(0)$ using the identical magnon-phonon coupling constants introduced in Fig. 4(e). Surprisingly, this simple calculation qualitatively captures the key features of the experimental $\Delta\kappa_{xx}(H)/\kappa_{xx}(0)$ with comparable magnitude: with increasing temperature, increasing $\kappa_{xx}$ turns into decreasing $\kappa_{xx}$ and the sharp suppression around the spin-flop transition fades out [left of Fig. 6]. Another remarkable point is that non-zero $\alpha_{\mathrm{ex}}$ is quite essential to reproduce the decreasing $\kappa_{xx}$ after the spin-flop transition around the $T_{\mathrm{N}}$ (see Fig. 6). In other words, the magnon-phonon hybridization induced by the exchange striction is crucial for longitudinal heat transport in MnPS$_3$ for $T<T_{\mathrm{N}}$.

However, this model cannot reproduce increasing $\kappa_{xx}$ data for $T>1.5T_{\mathrm{N}}$ as shown in Fig. 2(b), which is often found in the paramagnetic phase of several magnetic insulators [6,9,11,29,32,106]. Since magnon is no longer well-defined in the paramagnetic phase, it is hard to expect magnon-phonon hybridization. Hence, we can assume that the $\Delta\kappa_{xx}(H)/\kappa_{xx}(0)$ is now dominated by $\tau_{\lambda,\mathbf{k}}$ instead of $\left[E_{\lambda,\mathbf{k}}\frac{\partial\rho(E_{\lambda,\mathbf{k}})}{\partial T}\right]v_{\lambda,\mathbf{k}}^2$. A possible scenario is that the paramagnetic spin fluctuation can be a new scattering source [22,107,108], and the magnetic field suppresses the spin fluctuation, leading to an increase in $\kappa_{xx}$.

## V. SUMMARY

In summary, we have carried out thermal transport measurements on MnPS$_3$, a Néel-type 2D honeycomb antiferromagnet. The obtained $\kappa_{xx}$ shows distinct field behavior while $\kappa_{xy}$ displays a monotonic field dependence with a hump around the spin-flop transition. For the numerical analysis, we considered the realistic magnetic ground state of MnPS$_3$ and constructed the minimum model Hamiltonian, including both types of the spin-lattice coupling on equal footing: single-ion magnetostriction ($\mathcal{H}_{\mathrm{mp,SI}}$) and exchange striction ($\mathcal{H}_{\mathrm{mp,ex}}$) of Heisenberg interaction. Our LSWT calculation succeeded in capturing the key features in $\kappa_{xx}$ and $\kappa_{xy}$ only when including $\mathcal{H}_{\mathrm{mp,ex}}$ in addition to $\mathcal{H}_{\mathrm{mp,SI}}$. Our result suggests that $\mathcal{H}_{\mathrm{mp,ex}}$ should be considered on an equal footing with $\mathcal{H}_{\mathrm{mp,SI}}$ for a complete description of the magnon-phonon-driven THE.

## ACKNOWLEDGEMENTS

We thank Chaebin Kim, Seokhwan Yun, Stephen Winter, and Sang-Wook Cheong for the helpful discussions. The work at SNU is funded by the Leading Researcher Program of the National Research Foundation of Korea (Grant No. 2020R1A3B2079375). This work was also partly supported by the Ministry of Education through the core center program (2021R1A6C101B418). The work at KAIST is supported by the Brain Pool Plus Program through the National Research Foundation of Korea, funded by the Ministry of Science and ICT (2020H1D3A2A03099291) and the National Research Foundation of Korea, funded by the Korean Government via the SRC Center for Quantum Coherence in Condensed Matter (RS-2023-00207732). G.G. acknowledges support by the National Research Foundation of Korea (NRF-2022R1C1C2006578).

# FIGURES

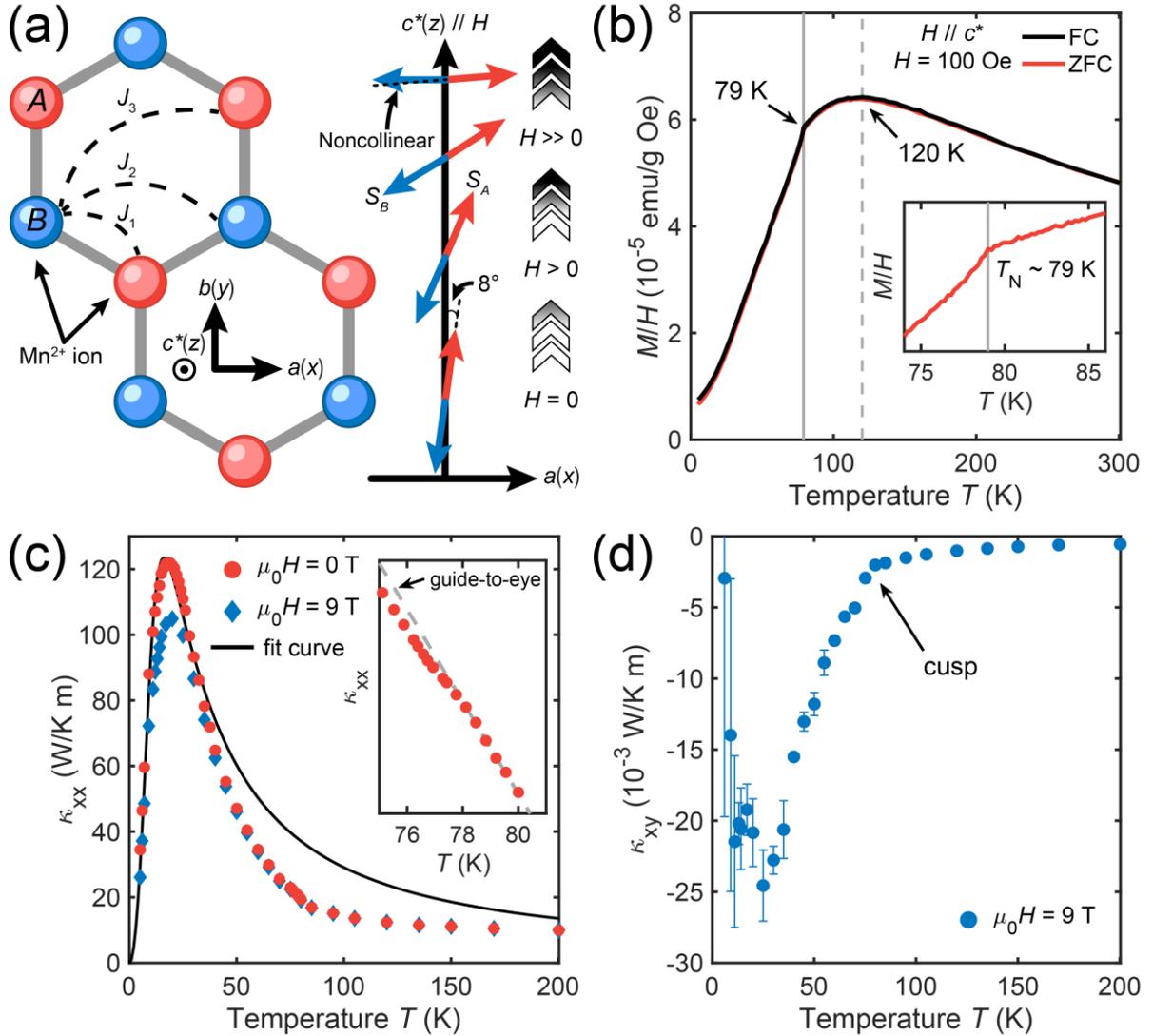

**FIG. 1.** (a) Left: Schematic of the $Mn^{2+}$ honeycomb layer in $MnPS_3$. $A$ and $B$ denote two sublattices of honeycomb lattice. $J_n$ is the exchange parameter of the $n$-th nearest-neighbor bond. Right: Schematic of the spin configuration under the different magnetic fields ($H$). The red (blue) arrow represents the magnetic moment of $Mn^{2+}$ of the $A(B)$-sublattice. (b) Magnetization ($M$) as a function of temperature. The 100 Oe of $H$-field was applied along the $c^*$-axis. FC and ZFC represent Field-Cooled and Zero-Field-Cooled, respectively. Gray solid- and dashed- lines indicate 79 K and 120 K, respectively. The inset shows an enlarged image around $T_N$. (c) $\kappa_{xx}$ as a function of temperature. Red circles and blue diamonds represent experimental data obtained from $\mu_0 H = 0$ T and 9 T, respectively. Black solid curve denotes the fitting result obtained from the Debye-Callaway model. Inset shows an enlarged image of $\kappa_{xx}$ around $T_N$. Gray dashed-line is a guide-to-eye specifying small anomalies around $T_N$. (d) $\kappa_{xy}$ as a function of temperature at $\mu_0 H = 9$ T. Error bars are standard deviations obtained from multiple measurements.

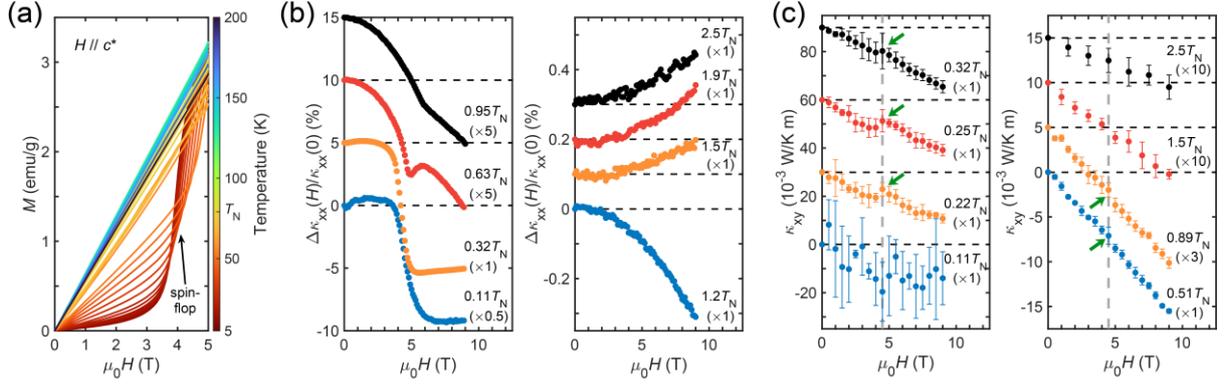

**FIG. 2.** (a) The magnetic field (*H*) dependence of magnetization (*M*) at several temperature points. (b) The magneto-thermal conductivity $\Delta\kappa_{xx}(H)/\kappa_{xx}(0)$ defined as $[\kappa_{xx}(H) - \kappa_{xx}(0)]/\kappa_{xx}(0)$ as a function of *H*-field. (c) The *H*-field dependence of $\kappa_{xy}$. The gray dashed-line is guide-to-eye specifying $\mu_0 H$ = 4.5 T. Green arrows indicate anomalies observed around the spin-flop transition. Error bars are standard deviations obtained from multiple measurements. For clarity of (b) and (c), the data are shifted upwards and multiplied by a scaling factor in parenthesis.

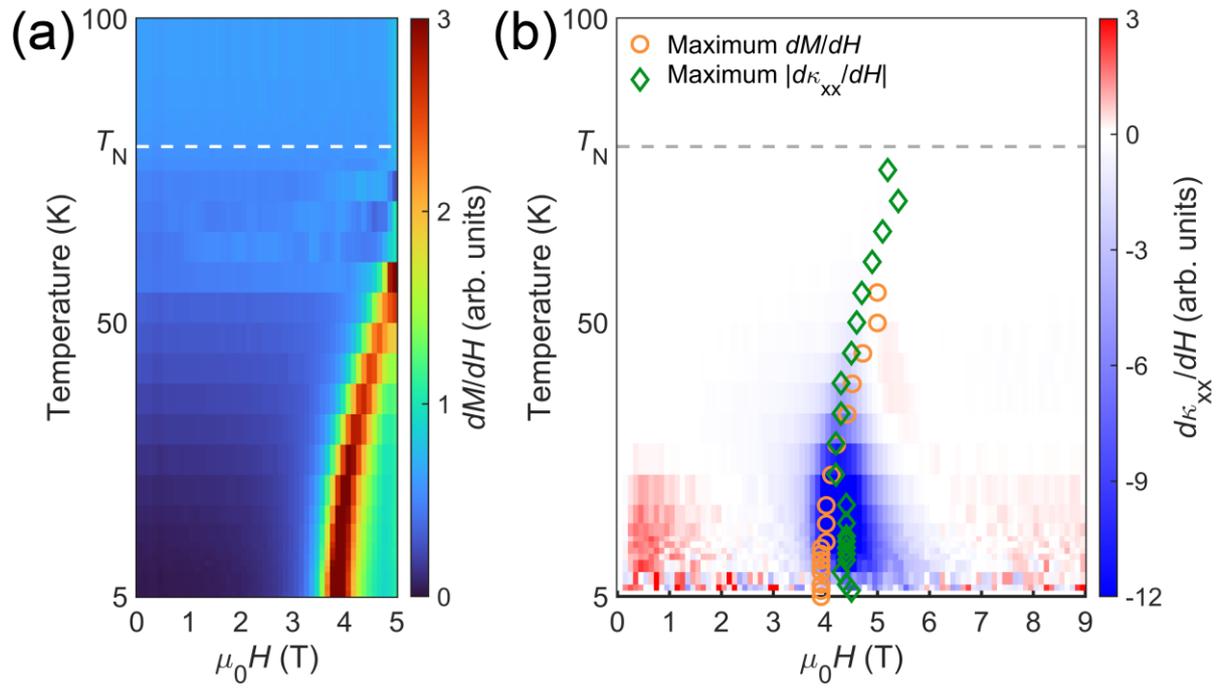

**FIG. 3.** The contour plots of (a) *dM/dH*, derivative of magnetization (*M*) with respect to *H*-field, and (b) $d\kappa_{xx}/dH$, $\kappa_{xx}$ differentiated with respect to *H*-field. The *x*- and *y*-axes denote *H*-field and temperature, respectively. Orange circles and green diamonds in (b) represent *H*-field values possessing maximum of *dM/dH* and $|d\kappa_{xx}/dH|$, respectively.

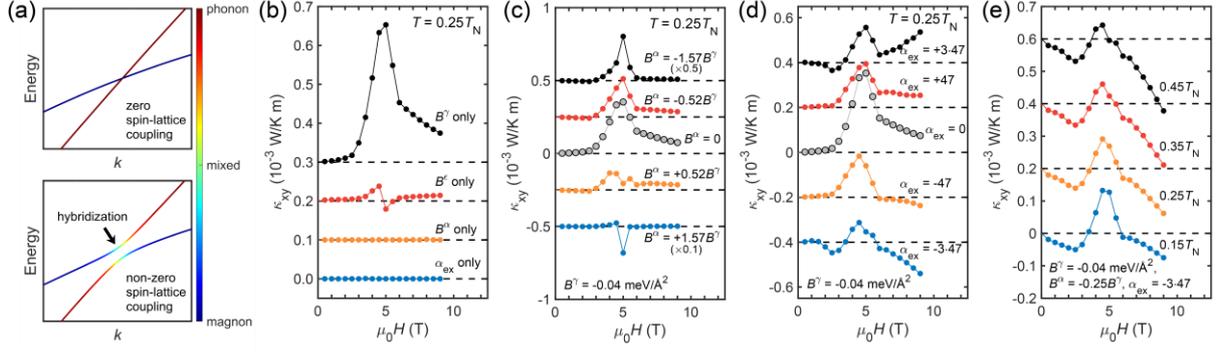

**FIG. 4.** (a) Schematic of magnon-phonon hybridization with respect to spin-lattice coupling. Red (blue) color denotes phonon (magnon)-like mode and green color represents hybridized mode. Upper panel: Magnon and phonon crossing point show no gap without spin-lattice coupling. Lower panel: Magnon and phonon crossing point are gapped by non-zero spin-lattice coupling. (b)-(e) Numerical calculation result of $\kappa_{xy}$ with the following spin-lattice coupling constants. (b) Each result obtained from $B^\gamma = -0.04$ meV/Å$^2$, $B^\epsilon = -0.028$ meV/Å$^2$, $B^\alpha = -0.01$ meV/Å$^2$, and $\alpha_{\text{ex}} = -47$, respectively. (c) $B^\gamma = -0.04$ meV/Å$^2$, $B^\alpha = -1.57B^\gamma, -0.52B^\gamma, +0.52B^\gamma, +1.57B^\gamma$, and $B^\epsilon = \alpha_{\text{ex}} = 0$. (d) $B^\gamma = -0.04$ meV/Å$^2$, $\alpha_{\text{ex}} = -3 \cdot 47, -47, +47, +3 \cdot 47$, and $B^\alpha = B^\epsilon = 0$. (e) $B^\gamma = -0.04$ meV/Å$^2$, $B^\alpha = -0.25B^\gamma$, $\alpha_{\text{ex}} = -3 \cdot 47$, and $B^\epsilon = 0$. For clarity, all the results in (b)-(e) are shifted upwards and multiplied by a scaling factor in parenthesis (multiplied unity if there is no parenthesis).

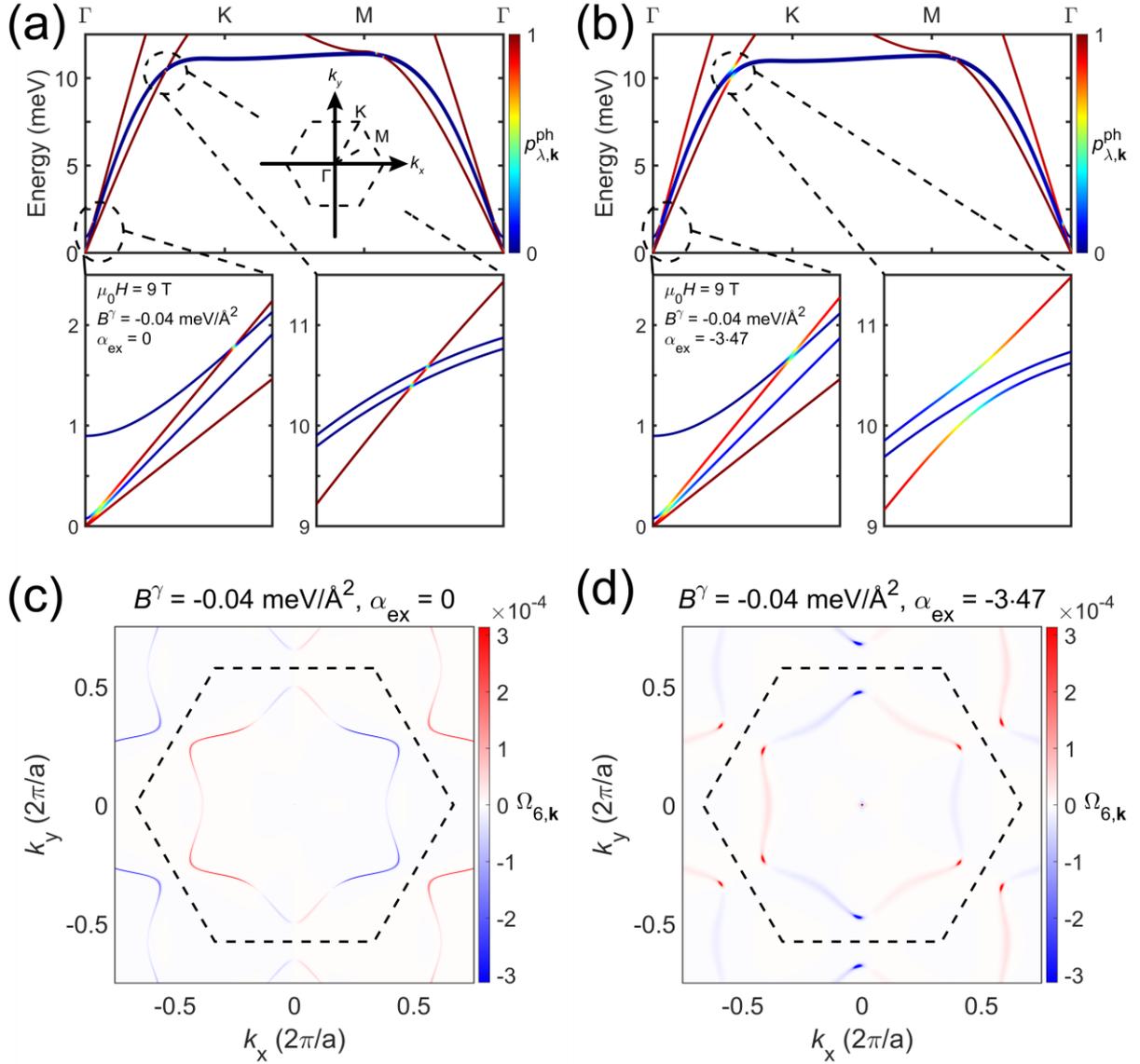

**FIG. 5.** The magnon and phonon band dispersions at $\mu_0 H = 9$ T obtained from (a) $B^\gamma = -0.04$ meV/Å$^2$, $\alpha_{ex} = 0$ and (b) $B^\gamma = -0.04$ meV/Å$^2$, $\alpha_{ex} = -3\cdot 47$. Insets show enlarged image near the $\Gamma$- and K-points. The resulting Berry curvature flux of the lowest band obtained from (c) $B^\gamma = -0.04$ meV/Å$^2$, $\alpha_{ex} = 0$ and (d) $B^\gamma = -0.04$ meV/Å$^2$, $\alpha_{ex} = -3\cdot 47$. Black dashed hexagon denotes the first Brillouin-zone boundary.

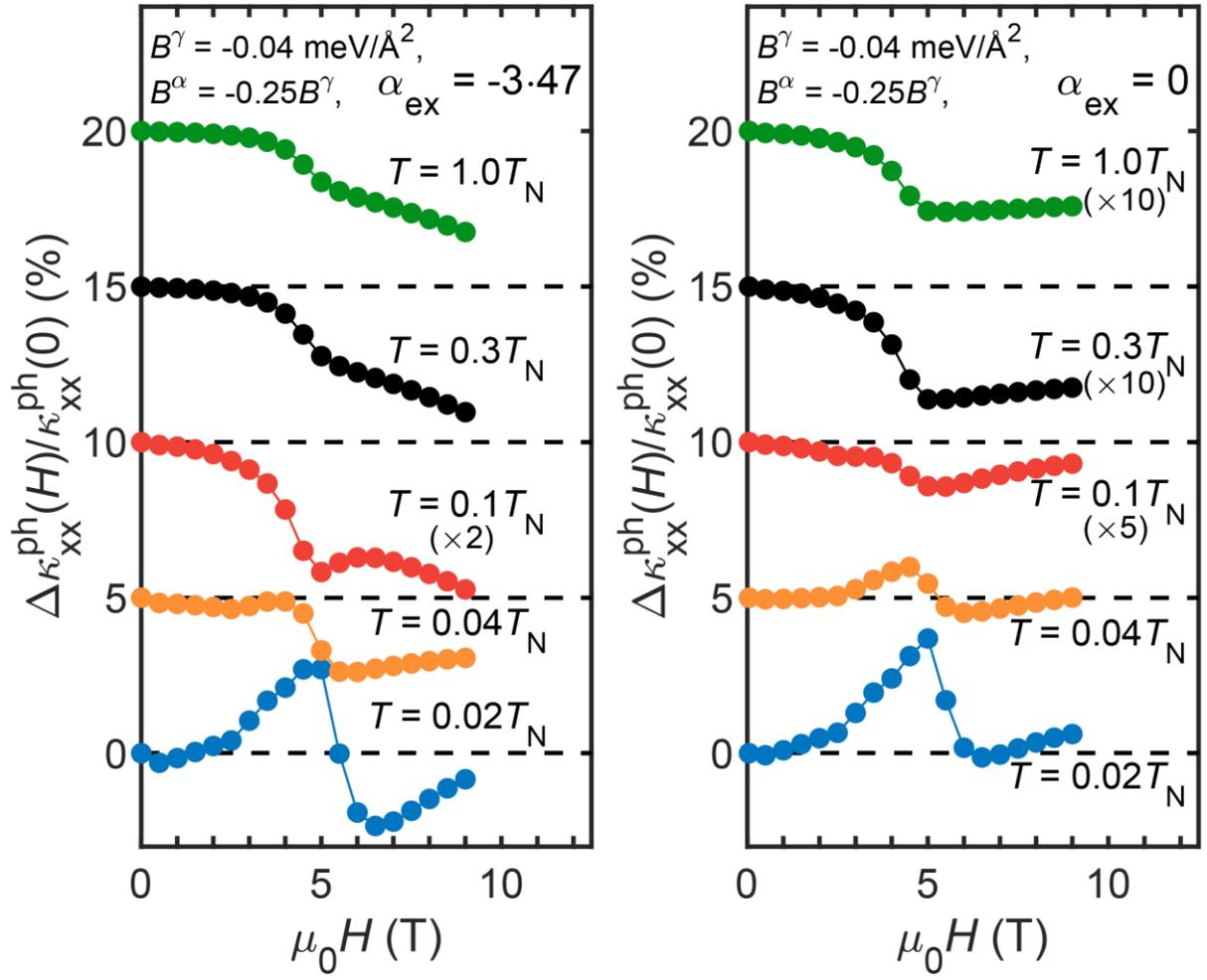

**FIG. 6.** Numerical calculation result of $\Delta\kappa_{xx}^{\text{ph}}(H)/\kappa_{xx}^{\text{ph}}(0)$ using the following parameter sets. Left: $B^{\gamma} = -0.04$ meV/Å$^2$, $B^{\alpha} = -0.25B^{\gamma}$, $\alpha_{\text{ex}} = -3 \cdot 47$. Right: $B^{\gamma} = -0.04$ meV/Å$^2$, $B^{\alpha} = -0.25B^{\gamma}$, $\alpha_{\text{ex}} = 0$. For clarity, the results are shifted upwards and multiplied by a scaling factor in parenthesis (multiplied unity if there is no parenthesis).